# Observation, Evidence and Origin of Room Temperature Magnetodielectric Effect in LaGa$_{0.4}$Mn$_{0.6}$O$_{3+\gamma}$

Hari Mohan Rai, Shailendra K. Saxena, Vikash Mishra, Rajesh Kumar and Pankaj R. Sagdeo[*]

Material Research Lab. (MRL), Department of Physics and MSE; Indian Institute of Technology Indore, Khandwa road, Simrol, Indore (M.P.) – 453552, India.

**Abstract**

We report an observation of room temperature magnetodielectric (RTMD) effect in LaGa$_{0.4}$Mn$_{0.6}$O$_{3+\gamma}$. Results of frequency dependent magnetoresistance (FDMR) measurements discards the possibility of any magnetoresistive contribution in the observed MD effect. The intrinsic nature of MD coupling has been validated/evidenced by means of magnetic field dependent Raman spectroscopy and explained in terms of modified volume strain governed by magnetic field induced rerotation of spin coupled Mn-orbitals. Ultimately, present RTMD effect is attributed to magneto-compression/magnetostriction associated with spin-phonon coupling as evidenced in the form of magnetic field induced hardening of symmetric stretching (*SS*) MnO$_6$ octahedral Raman modes. Presently studied Mn doped LaGaO$_3$ can be a candidate for magnetodielectric applications.

**Key words:** magnetodielectric, magnetoelectric, magnetocapacitance, multiferroics

**PACS:** 62.50.-p, 71.30.+h, 78.30.-j, 75.85.+t, 77.84.-s, 77.84.Dy, 75.80.+q

**Introduction**

Magnetoelectric (ME)/ magnetodielectric (MD) materials are of very high research interest due to their potential possible applications and interesting physics[1–14]. The ME effect i.e., intriguing coupling of magnetization (electric polarization) with external electric field (magnetic field), opens new possibilities in microelectronic storage systems. For example, making of ME-random access memory (ME-RAM) can tremendously improve its storage efficiency by reducing the time and power consumption required for 'reading and/or writing' operation of data[8]. In existing dynamic (DRAM) and magnetoresistive counterpart (i.e., (MRAM)), such 'read/write' operations are relatively very slow and more power consuming[8]. The MD phenomenon/terminology is observationally distinguishable from ME effect as it represents magnetic field induced change in capacitance i.e., dielectric permittivity ($\varepsilon'$), whereas, ME effect presents electric (magnetic) field induced change in magnetization (electric polarization). In literature, MD is termed also as magnetocapacitance (MC). Observation of MD effect has been reported in different single phase materials [15–19]. But, this effect is witnessed below room temperature (RT). For example - 3% (change in $\varepsilon'$) with H = 0.2 T below 30 K in $Tb_3Fe_5O_{12}$ [20], 13% at 0.5 T and 10 K in $Y_3Fe_5O_{12}$ [21], 1% in $CaMn_7O_{12}$ at 10 K [18], at 25 K in $EuMnO_3$ [22], at 230 K in CuO [23], at 28 K in $TbMnO_3$ [15], and also in other MD materials [6,24–26]. However, for MD effect to be used at device level, only MD materials exhibiting this effect near/at RT and with intrinsic nature are important. It is vital to note that observation of ME coupling needs recording the effect of an electric field on magnetization or, conversely, that of a magnetic field on ferroelectric polarization. A difficulty arises in both of these situations, as many of the contenders (especially perovskites having transition metal (TM) ions with mixed charge states) for being magnetoelectrics are indeed poor insulators, which makes it difficult for them to sustain such a high electric field necessary for realizing the effect i.e. to switch (align) the polarization (dipoles) [6,7,10,27–29]. By keeping this in view, a relatively simple and widely accepted alternative method is used; in which, dielectric constant ($\varepsilon'$) is measured as a function of applied magnetic field (H). The intrinsicality of MC observed through this indirect approach, is always a major concern, because, in many materials, it appears without any intrinsic ME coupling[1,6,7,30–34]. Generally, in such systems, MC arises due to extrinsic artifacts like magnetoresistance (MR)[1,6,7,30–34]. As far as intrinsic MD coupling is concern, ideally it can be realized in a material with magnetically switchable net electric dipole moment and such systems are barely formed [35]. Nonetheless, it can be achieved by making a compound which expands/shrinks in response to an externally applied magnetic field as the $\varepsilon'$ or ultimately capacitance $C$, is directly related to sample's dimensions by following relation-

$$C \propto \frac{A}{d} \qquad (1)$$

where, $C$ is the capacitance of a dielectric material, $A$ the area and $d$ is he thickness of dielectric. By keeping this alternative in view, presently, $LaGaO_3$ (LGO) is doped with Mn at Ga site with an expectation of RT (near RT) MD coupling because, the strain evolved due to significant difference in Ga-O (1.97 Å) and Mn-O (2.18Å long bond/1.90Å short bond) distances[36], is supposed to be magnetically switchable as the field transforms[36]/rerotates[13] Mn spin moments and hence the Mn-orbitals[13,36] (magnetic field induced re-rotation of spin coupled Mn-orbitals is possible[13,36], as in such systems spin and orbital degrees of freedoms are strongly coupled with each other). This magnetically switchable rerotation/retransformation of Mn-orbitals[13,36] may result into shrinking/expanding of material ($A/d$, in equation (1)) and hence, may result in an intrinsic MD effect. The composition $LaGa_{0.4}Mn_{0.6}O_3$ was prepared purposefully, because, across Ga:Mn ratio of 60:40/40:60, in $LaMn_{1-x}Ga_xO_3$, two different phases (orbitally, disordered *O-phase* and ordered *O'-phase*) appears within the *Pbnm* orthorhombic structure[37,38] and Lawes et al.[39] have suggested that significant MD response may appear near a phase change, therefore, a significant MD coupling can be expected in $LaGa_{0.4}Mn_{0.6}O_3$ sample.

## Experimental

For present study, Polycrystalline samples of $LaGa_{0.4}Mn_{0.6}O_3$ was prepared through conventional solid-state reaction route by following the process described in previous reports on Mn and Fe doped LGO [40,41]. The purity of structural phase is validated by using x-ray diffraction (XRD) experiments carried out in angle dispersive mode using Huber 5020 diffractometer at BL-12 XRD beam line on Indus-2 synchrotron radiation source. XRD data is refined by using Fullprof Rietveld refinement package[42,43]. For further measurements, these single phase powdered samples were pelletized at a pressure of 15 ton to form 1 mm thick circular discs of 12 mm diameter. These pellets were sintered in air at 1400 °C for 24 hours. Further, after being coated with silver paint these pellets were fired at 300° C for 30 min. For all as prepared pellets, RT capacitance and FDMR (i.e., magnetic field dependent change in real and imaginary parts of impedance) are measured in the absence and presence of magnetic field by means of a precision impedance analyzer. These measurements were performed by applying an oscillator voltage of ±1 V and the data was recorded for probing frequencies ranging from 30 Hz to 1MHz. The direction of applied magnetic field (H) was kept along the applied electric field to ensure the complete absence of contributions, which, may arise in MD measurements due to Hall-effect geometry. Oxygen stoichiometry, Mn charge state and $Mn^{3+}/Mn^{4+}$ ratio in $LaGa_{0.4}Mn_{0.6}O_3$ were estimated through iodometric titration method followed by calculations using charge neutrality condition (details are given in supplemental material (SM)). The magnetic field dependent RT Raman spectra on highly dense pellet were recorded from 300 to 1200 $cm^{-1}$ by using a Horiba Lab RAM high-resolution (~0.4 $cm^{-1}$) micro Raman spectrometer in confocal mode equipped with 30 mW air cooled He-Ne laser (632.8nm) as the excitation source and a charge-coupled device (CCD) detector. The spectra were recorded by using 50x standard objective with a spot size of ~5 μm whereas the laser power was kept at 15 mW.

## Results and discussions

Figure 1 shows the refined powder XRD pattern for polycrystalline sample of $LaGa_{0.4}Mn_{0.6}O_3$. This XRD pattern is refined with the help of Fullprof package by considering orthorhombic structure with *Pnma* space group [40]. The value for the goodness of fit (i.e., $\chi^2$) was found to be 1.46. The absence of any unaccounted peak in the refined XRD pattern, confirms the phase purity of prepared sample. In order to check the oxygen stoichiometry of presently studied $LaGa_{0.4}Mn_{0.6}O_3$ sample, iodometric titration experiments [44] have been carried out. The sample were found to be oxygen off-stoichiometric with some excess oxygen (i.e., coexistence of $Mn^{3+}$ and $Mn^{4+}$). In $LaGa_{0.4}Mn_{0.6}O_{3+\gamma}$, the value of γ is found to be 0.042. With this value of γ, the charge state of Mn and percentage of $Mn^{3+}$ and $Mn^{4+}$ in $LaGa_{0.4}Mn_{0.6}O_{3.042}$ (LGM6O) has been calculated by using the charge neutrality condition (SM). As a result 86% of total Mn is found to be present in 3+ state and remaining 14% is $Mn^{4+}$, whereas the charge state of Mn in LGM6O has been estimated as 3.14 (SM). The coexistence of $Mn^{3+}$ and $Mn^{4+}$ has been supported through XANES measurements (as the charge state determination by means of XANES is already being used [40,45–48]) carried out at Mn K-edge energy of 6539 eV. The value of Mn charge state and concentration of $Mn^{3+}/Mn^{4+}$ (for LGM6O) determined through titration, have been found closely consistent with XANES results. Note that, to further validate the oxygen excess, neutron diffraction measurements will be very useful as the oxygen excess in $LaMnO_{3.12}$ has already been studied by using neutron diffraction data [49,50] and explained in terms of cation vacancies [49,50].

Figure 2 shows the RT frequency dependence of capacitance and MC% (={($C$(H=2.35 T)-$C$(H=0 T))/( $C$(H=0 T)}x100) for LGM6O. This data is recorded in the absence (H=0 T) and presence (uoto H=2.35 T) of magnetic field. Similar, frequency dependent dielectric behaviour under zero magnetic field, has already been discussed in detail elsewhere[40] for $LaGa_{1-x}Mn_xO_3$ (x=0, 0.05, 0.1, 0.2 and 0.3). As far as present magnetic field dependent data is concern, the $C$ is found to be decreased over the entire range of probing frequencies due to the application of magnetic field (H=0T to H=2.35T) as depicted in Figure 2 and its inset. Further, the MD effect is expected to be MR free or intrinsic when it is measured at probing frequencies corresponding to grains contribution (generally at $\geq$ 100 kHz) [7,40,51]. Since, at such high probing frequencies, the hopping charge transport (common in perovskites having Mn ions with mixed valence) gets supressed as the carriers have no chance to jump between hopping cites[7,40,51] (presently $Mn^{3+}$ and $Mn^{4+}$), therefore, only grain contribution is expected over these high frequencies. This is the reason, the MD data recorded around 100kHz or higher frequencies, is generally emphasized [29,52–54] for MD analysis. Presently, a significant RTMD effect (i.e., MC%) of about -3% is observed in LGM6O sample when measured at a probing frequency of 1MHz (Figure 2). Since MC% is negative, presently observed MD effect can be MR affected only with positive MR[1,7,55]. The magnitude of MC% (red curve in Figure 2) is comparable with the MD effect reported by others at same/lower probing frequency in different materials[18,53,55–57], but, with the application of equivalent[57]/relatively higher magnetic fields [18,53,55–57]. It should be noted that the presently observed MC% is well above the sum of instrumental and statistical error bar, even at the highest probing frequency (1MHz). As far as magnetic field dependent dielectric loss (tan$\delta$) of LGM6O is concern, opposite to $C$ ($\varepsilon'$), *tan$\delta$* increases over the entire range of probing frequency due to the application of magnetic field as shown in the main panel and corresponding insets of Figure S1(SM). The application of magnetic field makes electron hopping (via double exchange) easier, between $Mn^{3+}$ and $Mn^{4+}$, consequently, '$Mn^{3+}/Mn^{4+}$' dipoles start oscillating with a higher frequency[40] and hence low frequency tan$\delta$-anomaly corresponding to this characteristic frequency is observed to be shifted towards higher frequency side due to the application of magnetic field (Figure S1 of SM). Further, as this magnetic field mediated hopping transport of electron can increase conductivity $\sigma$, hence, according to tan$\delta$=($\omega\varepsilon''$+ $\sigma$)/$\omega\varepsilon'$, the loss tangent is increased over the entire range of probing frequencies (Figure S1 and corresponding insets in SM) due to finite magnetic field. The trend of magnetic field dependent MC% and tan$\delta$ are opposite and indicate a positive and negative dc MR [1,7,55], providing contradictory information. However, as the MD measurements are performed by applying an electric field with varying frequency upto MHz or even higher range[1,6,8,29–34,54,55,58–62]. Thus, to get a better insight about the contribution of MR in the presently observed MD effect, FDMR (20 Hz to 1 MHz) has been analysed by means of MRIS (MR via impedance spectroscopy) measurements. These RT measurements are performed in the absence (H=0.0 T) and presence of magnetic field (uoto H=2.35 T) to record the real ($Z'$) and imaginary ($Z''$) part of impedance. The data of these MRIS measurements is shown as a Cole-Cole (Nyquist) plot [40] in Figure 3. It is clear that at frequencies $\leq 10^5$Hz, the resistance ($Z'$) and reactance ($Z''$), both are decreasing significantly due to external magnetic field (upto H=2.35T) which signifies negative FDMR over these probing frequencies. Whereas, at frequencies $\geq 10^5$Hz, neither part of impedance is changing significantly due to the application of external magnetic field (upto H=2.35T). It is important to note that this negative FDMR is not affecting the negative MD effect observed in LGM6O (Figure 2), because, any kind (dc or ac) of negative MR, leads to a positive MD response[1,7,55]. Thus, it seems that the MD coupling in LGM6O (Figure 2) is completely MR unaffected and very strong too as it is appearing by suppressing the effect of negative FDMR even at low frequencies. This is the reason, the identification of different segments (corresponds to space charge polarization (SCP), grain boundary (GB) and grains)[63] in case of LGM6O was not required and hence the corresponding $Z'$-$Z''$ data has not been fitted. Ultimately, unlike to the consideration of trends convention by Mao *et al.*[55], even with capacitance and *tanδ* changing oppositely due to magnetic field, LGM6O exhibits RTMD effect without any resistive contribution over the entire range of probing frequencies. Thus, unlike to the consideration by Mao *et al.*[55], we propose that only trends of magnetic field dependent changes in capacitance (or ε′) and *tanδ* are not sufficient to correctly estimate/analyse the contribution of MR in MC%. Further, the MD response is recorded through electrical measurements, therefore appearance of FDMR at low frequencies (though it is not affecting MD coupling) is understandable. However, magnetic field dependent measurements using non-electrical characterization techniques (where, electrical resistance does not contribute in signal) may provide additional and concrete information about the intrinsic nature and mechanism of MD coupling in presently studied sample. Thus, magnetic field dependent Raman spectra for LGM6O were recorded between 300 to 1200 cm$^{-1}$ and the results are shown in Figure 4. Similar to most of the manganitelike $ABO_3$ systems, the Raman spectrum is basically dominated by the $MnO_6$ octahedral vibrational modes[37,38,64,65]. Using previous conventions, the broad features observed around 500 cm$^{-1}$ and above 600 cm$^{-1}$ (Figure 4) were respectively assigned as bending (*B*) or Jahn-teller asymmetric stretching (*JT-AS*) and symmetric stretching (*SS*) MnO$_6$ octahedral Raman modes[37,38,66,67]. Interestingly, with the application of a very low magnetic field of ≤0.6 T, the peak position of *SS* mode is found to be significantly shifted toward higher frequency side (Figure 4). A shift of 5.1 cm$^{-1}$ between the peak positions corresponding to zero field (634.7 cm$^{-1}$) and lowest applied field i.e., H= 0.35 (639.8 cm$^{-1}$) has been marked with the help of a vertical line and two arrows (Figure 4) whereas a forward shift of 7.2 cm$^{-1}$ is observed for highest available field of H=0.6 T(641.9 cm$^{-1}$). This magnetic field dependent forward shifting is a signature of hardening of *SS* MnO$_6$ octahedral Raman modes, indicating compression of material [37] due to magnetic field. As a result of this volume compression (considering uniform compression for circular pellets), presently observed decrease in capacitance i.e., negative MC%, appears justified in accordance to relation (1) as in case of a circular pellet (sample) *C* scales with *A* (for uniform magnetic compression/expansion). As far as the origin of presently observed significant negative MC% or magnetic hardening of Raman modes (or magnetic compression) is concern, it can be understood as follows. In present LGM6O system 86% of total Mn is in its 3+ state which is JT distorted [40,68] due to that Mn$^{3+}$ forms two long Mn-O bonds (2.18 Å) [69] and four short Mn-O bonds (1.90 Å)[69] with octahedrally connected oxygens. Further, the ionic radius of Ga$^{3+}$ (0.62 Å) is a little smaller than that of the Mn$^{3+}$ (0.64Å) [69–71] whereas Ga-O distances (1.97 Å) is slightly higher than Mn-O short bond (1.90 Å) and significantly smaller to Mn-O long bond (2.18 Å)[69]. The evolution of orthorhombic strain in LaMn$_{1-x}$Ga$_x$O$_3$ is already predicted by Farrel et al.[69] in terms of transformation/rotation of Mn - $(3x^2 - r^2)/(3y^2 - r^2)$ orbital into the $(3z^2 - r^2)$ state due to the replacement of Mn$^{3+}$ with Ga$^{3+}$. The evolution of such orthorhombic strain can be considered in present LGM6O sample, by means of alignment of Mn orbitals transforming due to the replacement of non-JT active Ga$^{3+}$ ions with JT distorted Mn$^{3+}$ ions[69]. Further, as the application of magnetic field (upto H=2.35T) aligns/rotates magnetic moments related to Mn ions, therefore, readjustment/rerotaion[13] of already transformed Mn orbitals appears possible (as for such correlated systems, spin and orbital degrees of freedoms are strongly coupled) which may lead to a change in the above discussed orthorhombic strain (i.e., compression or expansion of material; presently compression) and hence in capacitance (or ε′) according to relation (1).

## Conclusions

In summary, an observation of RTMD effect has been reported in LGM6O. Over the entire range of probing frequencies (30 Hz to 1 MHz), the observed MD coupling is completely MR unaffected and very strong too as it appeared by suppressing the effect of negative FDMR even at low frequencies. The intrinsic nature of negative MC% has been evidenced in terms of magnetic field dependent hardening of *SS* MnO$_6$ octahedral Raman modes (i.e., magneto-compression). Present analysis reveals that only trends of magnetic field dependent changes in capacitance (or ε′) and *tanδ* are not sufficient to correctly estimate/analyse the contribution of MR in MC%. The rerotation of spin coupled Mn-orbitals under the influence of magnetic field was proposed as the mechanism responsible for the compression of material and ultimately for present RTMD effect. As a final remark, we propose that the intrinsic MD coupling can be realized in strongly correlated compounds that shrink/expand in response to an externally applied magnetic field.

**Acknowledgments**

The authors sincerely thank Prof. P. Mathur, Director of IIT Indore, for his encouragement. Authors are thankful to Dr. K. V. Adarsh (Associate professor, Indian Institute of Science Education and Research, Bhopal) and Mr. Rituraj Sharma (Research scholar, Indian Institute of Science Education and Research, Bhopal) for providing Raman facility and helping in Raman measurements. The authors sincerely thank Dr A. K. Sinha, Dr Archna Sagdeo, Mr. Anuj Upadhyay and Mr. Manvendra N. Singh for their help during x-ray diffraction measurements. CSIR, New Delhi is acknowledged for funding high temperature furnace under the project 03(1274)/13/EMR-II used for the sample preparations. DAE-BRNS is acknowledged for funding Impedance Analyzer used for the present measurements under the grant No. 2013/37P/31/BRNS. One of the authors (HMR) acknowledges the Ministry of Human Resource Development (MHRD), government of India for providing financial support as Teaching Assistantship.

**References**

[1] T. Bonaedy, Y.S. Koo, K.D. Sung, and J.H. Jung, Appl. Phys. Lett. **91**, 132901 (2007).
[2] V.G. Nair, A. Das, V. Subramanian, and P.N. Santhosh, J. Appl. Phys. **113**, 213907 (2013).
[3] V.G. Nair, L. Pal, V. Subramanian, and P.N. Santhosh, J. Appl. Phys. **115**, 17D728 (2014).
[4] V. Castel and C. Brosseau, Appl. Phys. Lett. **92**, 233110 (2008).
[5] V. Castel, C. Brosseau, and J.B. Youssef, J. Appl. Phys. **106**, 64312 (2009).
[6] W. Eerenstein, N.D. Mathur, and J.F. Scott, Nature **442**, 759 (2006).
[7] G. Catalan, Appl. Phys. Lett. **88**, 102902 (2006).
[8] K.F. Wang, J.-M. Liu, and Z.F. Ren, Adv. Phys. **58**, 321 (2009).
[9] M. Fiebig, J. Phys. Appl. Phys. **38**, R123 (2005).
[10] R. Ramesh and N.A. Spaldin, Nat. Mater. **6**, 21 (2007).

[11] N.A. Spaldin and M. Fiebig, Science **309**, 391 (2005).

[12] D. Khomskii, Physics **2**, 20 (2009).

[13] N. Imamura, M. Karppinen, T. Motohashi, D. Fu, M. Itoh, and H. Yamauchi, J. Am. Chem. Soc. **130**, 14948 (2008).

[14] C.N.R. Rao, A. Sundaresan, and R. Saha, J. Phys. Chem. Lett. **3**, 2237 (2012).

[15] T. Kimura, T. Goto, H. Shintani, K. Ishizaka, T. Arima, and Y. Tokura, Nature **426**, 55 (2003).

[16] N. Hur, S. Park, P.A. Sharma, J.S. Ahn, S. Guha, and S.-W. Cheong, Nature **429**, 392 (2004).

[17] B. Lorenz, Y.Q. Wang, Y.Y. Sun, and C.W. Chu, Phys. Rev. B **70**, 212412 (2004).

[18] M. Sánchez-Andújar, S. Yáñez-Vilar, N. Biskup, S. Castro-García, J. Mira, J. Rivas, and M.A. Señarís-Rodríguez, J. Magn. Magn. Mater. **321**, 1739 (2009).

[19] S. Weber, Phys. Rev. Lett. **96**, 15702 (2006).

[20] N. Hur, S. Park, S. Guha, A. Borissov, V. Kiryukhin, and S.-W. Cheong, Appl. Phys. Lett. **87**, 42901 (2005).

[21] Y. Yamasaki, Y. Kohara, and Y. Tokura, Phys. Rev. B **80**, 140412 (2009).

[22] J. Hemberger, Phys. Rev. B **75**, 35118 (2007).

[23] T. Kimura, Y. Sekio, H. Nakamura, T. Siegrist, and A.P. Ramirez, Nat. Mater. **7**, 291 (2008).

[24] R.S. 44, J.F. Mitchell, and P. Schiffer, Phys. Rev. B **72**, 144429 (2005).

[25] T. Kimura, S. Kawamoto, I. Yamada, M. Azuma, M. Takano, and Y. Tokura, Phys. Rev. B **67**, 180401 (2003).

[26] T. Suzuki and T. Katsufuji, Phys. Rev. B **77**, 220402 (2008).

[27] W. Eerenstein, F.D. Morrison, J.F. Scott, and N.D. Mathur, Appl. Phys. Lett. **87**, 101906 (2005).

[28] W. Eerenstein, F.D. Morrison, J. Dho, M.G. Blamire, J.F. Scott, and N.D. Mathur, Science **307**, 1203 (2005).

[29] P. Mandal, V.S. Bhadram, Y. Sundarayya, C. Narayana, A. Sundaresan, and C.N.R. Rao, Phys. Rev. Lett. **107**, 137202 (2011).

[30] K.D. Chandrasekhar, A.K. Das, C. Mitra, and A. Venimadhav, J. Phys. Condens. Matter **24**, 495901 (2012).

[31] K.D. Chandrasekhar, A.K. Das, and A. Venimadhav, J. Phys. Condens. Matter **24**, 376003 (2012).

[32] S. Ghara, K. Yoo, K.H. Kim, and A. Sundaresan, J. Appl. Phys. **118**, 164103 (2015).

[33] Z.X. Cheng, H. Shen, J.Y. Xu, P. Liu, S.J. Zhang, J.L. Wang, X.L. Wang, and S.X. Dou, J. Appl. Phys. **111**, 34103 (2012).

[34] A. Venimadhav, D. Chandrasekar, and J.K. Murthy, Appl. Nanosci. **3**, 25 (2012).

[35] N.A. Hill, J. Phys. Chem. B **104**, 6694 (2000).

[36] J. Farrell and G.A. Gehring, New J. Phys. **6**, 168 (2004).

[37] M. Baldini, D. Di Castro, M. Cestelli-Guidi, J. Garcia, and P. Postorino, Phys. Rev. B **80**, 45123 (2009).

[38] L. Malavasi, M. Baldini, D. di Castro, A. Nucara, W. Crichton, M. Mezouar, J. Blasco, and P. Postorino, J. Mater. Chem. **20**, 1304 (2010).

[39] G. Lawes, A.P. Ramirez, C.M. Varma, and M.A. Subramanian, Phys. Rev. Lett. **91**, 257208 (2003).

[40] H.M. Rai, S.K. Saxena, R. Late, V. Mishra, P. Rajput, A. Sagdeo, R. Kumar, and P.R. Sagdeo, RSC Adv. **6**, 26621 (2016).

[41] P. Singh, I. Choudhuri, H.M. Rai, V. Mishra, R. Kumar, B. Pathak, A. Sagdeo, and P.R. Sagdeo, RSC Adv. **6**, 100230 (2016).

[42] H. Rietveld, J. Appl. Crystallogr. **2**, 65 (1969).

[43] D.B. Wiles and R.A. Young, J. Appl. Crystallogr. **14**, 149 (1981).

[44] A. Shahee, R.J. Choudhary, R. Rawat, A.M. Awasthi, and N.P. Lalla, Solid State Commun. **177**, 84 (2014).

[45] A. Sagdeo, K. Gautam, P.R. Sagdeo, M.N. Singh, S.M. Gupta, A.K. Nigam, R. Rawat, A.K. Sinha, H. Ghosh, T. Ganguli, and A. Chakrabarti, Appl. Phys. Lett. **105**, 42906 (2014).

[46] C. Li, X. Han, F. Cheng, Y. Hu, C. Chen, and J. Chen, Nat. Commun. **6**, 7345 (2015).

[47] M. Sikora, C. Kapusta, K. Knížek, Z. Jirák, C. Autret, M. Borowiec, C.J. Oates, V. Procházka, D. Rybicki, and D. Zajac, Phys. Rev. B **73**, 94426 (2006).

[48] E.E. Alp, G.L. Goodman, L. Soderholm, S.M. Mini, M. Ramanathan, G.K. Shenoy, and A.S. Bommannavar, J. Phys. Condens. Matter **1**, 6463 (1989).

[49] M.A. Peña and J.L.G. Fierro, Chem. Rev. **101**, 1981 (2001).

[50] B.C. Tofield and W.R. Scott, J. Solid State Chem. **10**, 183 (1974).

[51] M.A. El Hiti, J. Magn. Magn. Mater. **192**, 305 (1999).

[52] K. Manna, R.S. Joshi, S. Elizabeth, and P.S.A. Kumar, Appl. Phys. Lett. **104**, 202905 (2014).

[53] Y. Kitagawa, Y. Hiraoka, T. Honda, T. Ishikura, H. Nakamura, and T. Kimura, Nat. Mater. **9**, 797 (2010).

[54] J.K. Murthy, K.D. Chandrasekhar, S. Murugavel, and A. Venimadhav, J Mater Chem C **3**, 836 (2015).

[55] X. Mao, H. Sun, W. Wang, X. Chen, and Y. Lu, Appl. Phys. Lett. **102**, 72904 (2013).

[56] C.-H. Yang, S.-H. Lee, T.Y. Koo, and Y.H. Jeong, Phys. Rev. B **75**, 140104 (2007).

[57] X. Wu, X. Wang, Y. Liu, W. Cai, S. Peng, F. Huang, X. Lu, F. Yan, and J. Zhu, Appl. Phys. Lett. **95**, 182903 (2009).

[58] G. Catalan and J.F. Scott, Adv. Mater. **21**, 2463 (2009).

[59] J.F. Scott, Science **315**, 954 (2007).

[60] A. Singh, V. Pandey, R.K. Kotnala, and D. Pandey, Phys. Rev. Lett. **101**, 247602 (2008).

[61] K. Everschor-Sitte, M. Sitte, and A.H. MacDonald, Phys. Rev. B **92**, 245118 (2015).

[62] R.A. Mondal, B.S. Murty, and V.R.K. Murthy, Phys.Chem.Chem.Phys. **17**, 2432 (2014).

[63] H.M. Rai, S.K. Saxena, V. Mishra, A. Sagdeo, P. Rajput, R. Kumar, and P.R. Sagdeo, J Mater Chem C **4**, 10876 (2016).

[64] A. Congeduti, P. Postorino, E. Caramagno, M. Nardone, A. Kumar, and D.D. Sarma, Phys. Rev. Lett. **86**, 1251 (2001).

[65] P. Postorino, A. Congeduti, E. Degiorgi, J.P. Itié, and P. Munsch, Phys. Rev. B **65**, 224102 (2002).

[66] I. Loa, P. Adler, A. Grzechnik, K. Syassen, U. Schwarz, M. Hanfland, G.K. Rozenberg, P. Gorodetsky, and M.P. Pasternak, Phys. Rev. Lett. **87**, 125501 (2001).

[67] M.N. Iliev, M.V. Abrashev, H.-G. Lee, V.N. Popov, Y.Y. Sun, C. Thomsen, R.L. Meng, and C.W. Chu, Phys. Rev. B **57**, 2872 (1998).

[68] J.B. Goodenough, A. Wold, R.J. Arnott, and N. Menyuk, Phys. Rev. **124**, 373 (1961).

[69] J. Farrell and G.A. Gehring, New J. Phys. **6**, 168 (2004).

[70] R.D. Shannon, Acta Crystallogr. Sect. A **32**, 751 (1976).

[71] J. Blasco, J. García, J. Campo, M.C. Sánchez, and G. Subías, Phys. Rev. B **66**, 174431 (2002).

# Figures

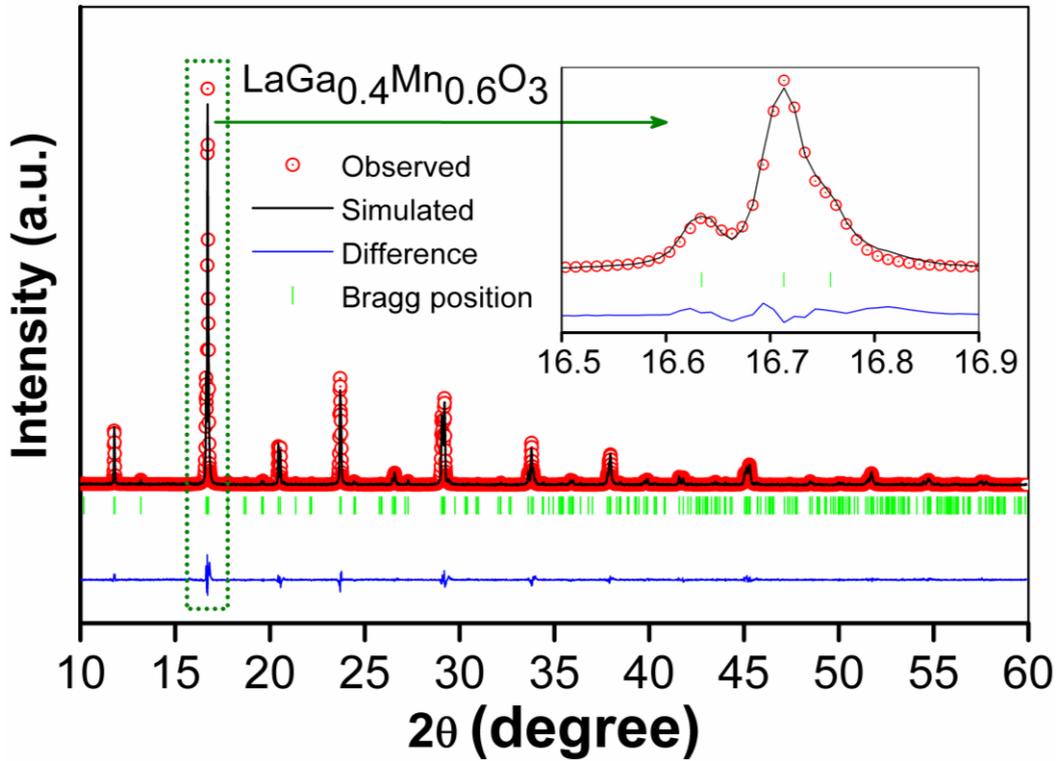

**Figure-1**

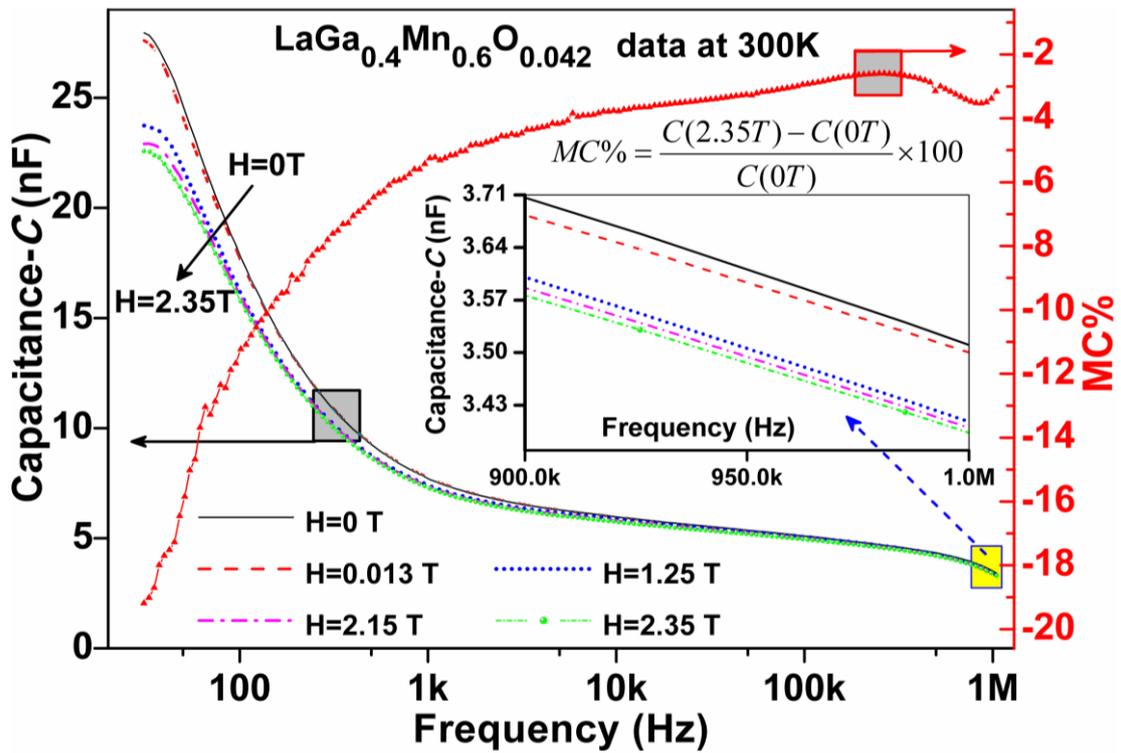

**Figure-2**

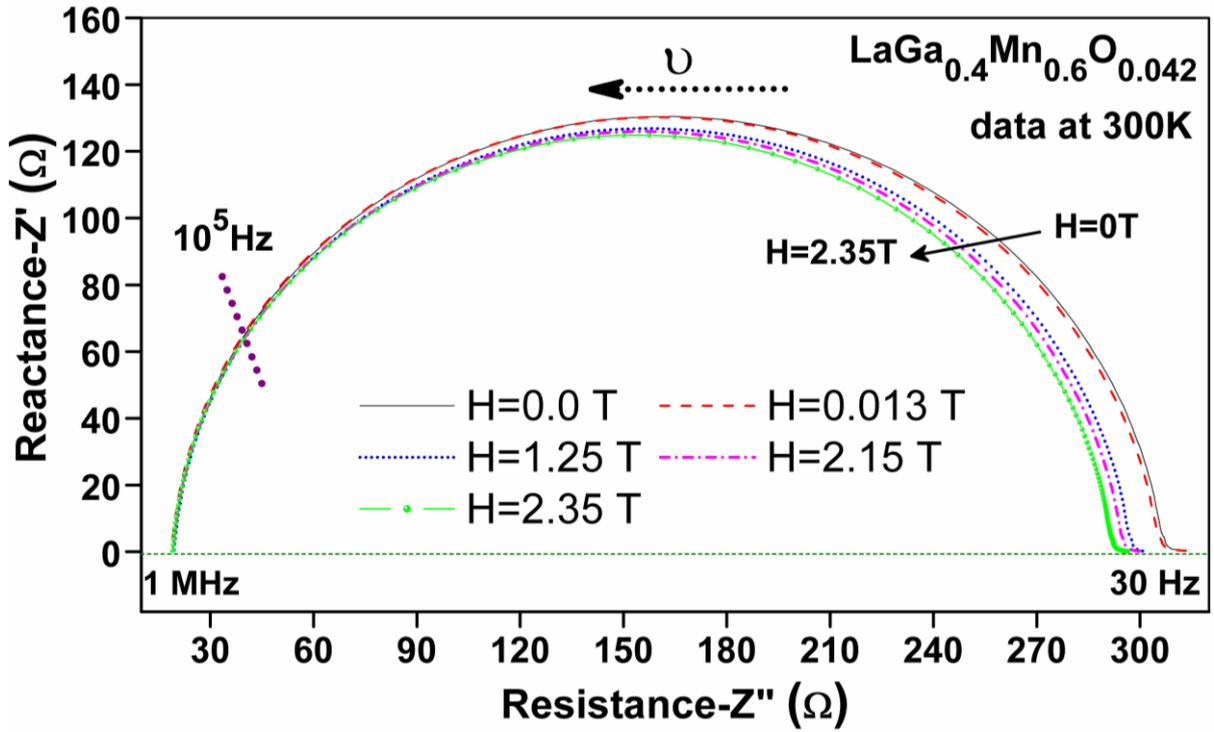

**Figure-3**

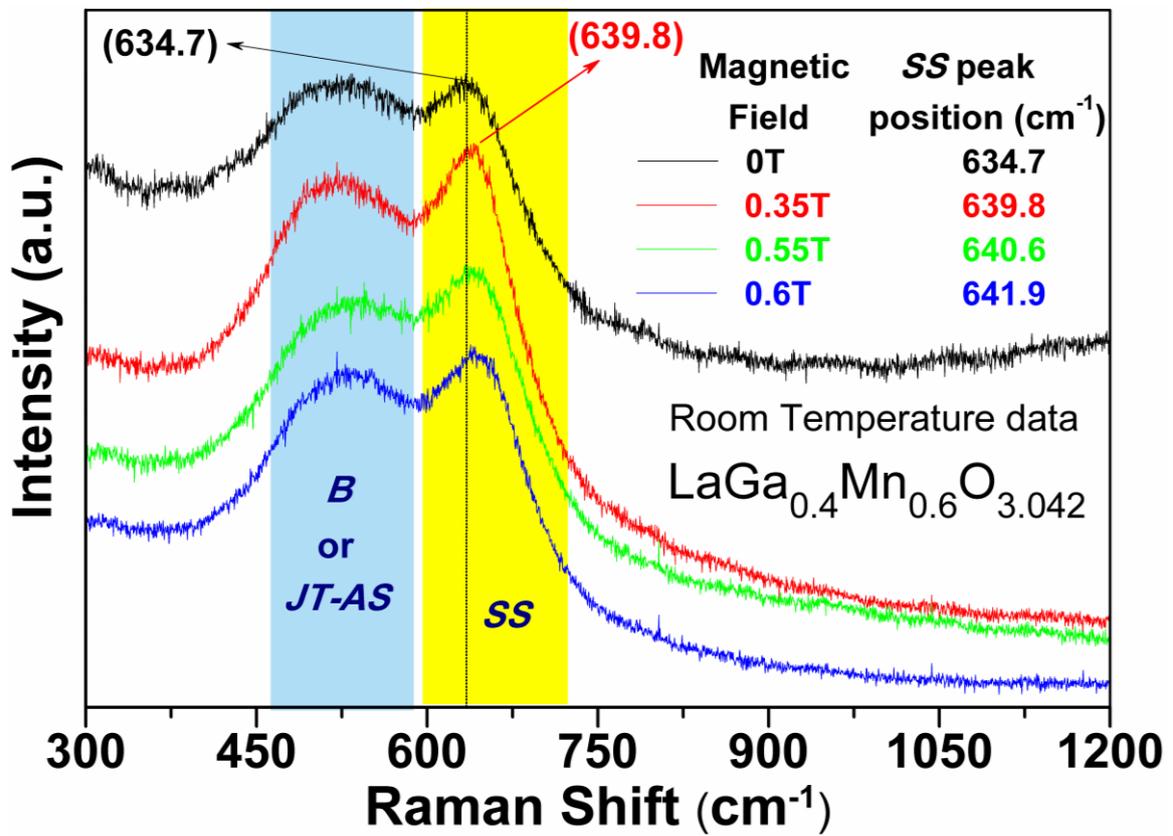

**Figure-4**

# Figure Captions

FIGURE-1: Rietveld refined X-ray diffraction data of LGM6O powder sample recorded at the Indus-2 synchrotron radiation source. Inset shows magnified view of most intense peak; reflecting the quality of fitting.

FIGURE-2: Room-temperature capacitance (measured in the absence (H=0T) and presence (upto H=2.35T) of magnetic field) and magnetocapacitance (MC%) as a function of frequency. Inset shows the magnified view of data across 1MHz. MC% has been calculated by using the formula shown in the Figure.

FIGURE-3: Room-temperature *Cole-Cole* diagram of LGM6O, plotted for frequencies ranging from 30 Hz to 1 MHz. The data is recorded in the absence (H=0T) and presence (upto H=2.35T) of magnetic field.

FIGURE-4: Room-temperature Raman spectra of LGM6O, recorded in the absence (H=0T) and presence (upto H=0.6T) of magnetic field. A vertical dotted line is drawn to show the shifting of *SS* $MnO_6$ octahedral Raman mode due to the application of magnetic field. The x-axis intersect of this dotted line represents the peak position (634.7 $cm^{-1}$) of *SS* $MnO_6$ octahedral Raman mode under zero magnetic field. The peak position of this *SS* mode for H=0T to H=0.6T are tabulated.

# Supplemental Material

## Observation, Evidence and Origin of Room Temperature Magnetodielectric Effect in LaGa$_{0.4}$Mn$_{0.6}$O$_{3+\gamma}$

Hari Mohan Rai[*], Shailendra K. Saxena, Vikash Mishra, Rajesh Kumar and Pankaj R. Sagdeo

Material Research Lab. (MRL), Department of Physics and MSE; Indian Institute of Technology Indore, Khandwa road, Simrol, Indore (M.P.) – 453552, India.

### (1) Estimation of oxygen content and multi-valence of Mn ions in LaGa$_{0.4}$Mn$_{0.6}$O$_3$

In order to find out the status of oxygen stoichiometry in presently studied LaGa$_{0.4}$Mn$_{0.6}$O$_3$ compound, iodometric titration experiment has been carried out. For this purpose sodium thio-sulphate (Na$_2$S$_2$O$_3$.5H$_2$O) solution is used as titrant whereas the analyte (LaGa$_{0.4}$Mn$_{0.6}$O$_3$) is dissolved in HCl solution and KI is added in that solution to get an end point as a result of redox. The titrant (in Burette) is added drop by drop in the Analyte+HCL+KI solution until it becomes colorless (end point). The end point is confirmed by using starch as an indicator. The volume of used titrant is recorded. Presently studied LaGa$_{0.4}$Mn$_{0.6}$O$_3$ compound was found to be oxygen off-stoichiometric with an excess of oxygen i.e., LaGa$_{0.4}$Mn$_{0.6}$O$_{3+\gamma}$, because gamma ($\gamma$) has been found to be positive as calculated by utilizing the well-known equation of normality (N$_1$V$_1$=N$_2$V$_2$). The value of $\gamma$ in LaGa$_{0.4}$Mn$_{0.6}$O$_{3+\gamma}$, has been estimated as 0.042. By using the value of $\gamma$, the charge state of Mn and percentage of Mn$^{3+}$ and Mn$^{4+}$ in LaGa$_{0.4}$Mn$_{0.6}$O$_{3.042}$ (LGM6O) has been calculated, in following manner, by using the charge neutrality condition.

**(i) Mn Charge state (say x)**

For the present sample, by charge neutrality condition,

$$La^{3+}Ga^{3+}_{0.4}Mn^{x}_{0.6}O^{2-}_{(3+\gamma)} = La^{3+}Ga^{3+}_{0.4}Mn^{x}_{0.6}O^{2-}_{(3+0.042)} = 0$$

Here, $x$ is the charge state of Mn in present sample and $\gamma = 0.042$ (by titration),

Thus,

$(3\times 1)+(3\times 0.4)+0.6x+(-2\times 3.042)=0$

$3+1.2+0.6x-6.084=0$

$4.2+0.6x-6.084=0$

$0.6x - 1.884 = 0$

$x = \dfrac{1.884}{0.6}$

$x = 3.14$

Thus, through iodometric titration, the charge state of Mn in present sample is found 3.14.

### (ii) Percentage of $Mn^{3+}$ and $Mn^{4+}$

For the present sample, by charge neutrality condition,

$$LaGa_{0.4}Mn_{0.6}O_{(3\pm\gamma)} = La^{3+}Ga^{3+}_{0.4}Mn^{3+}_{0.6(1-x)}Mn^{4+}_{0.6x}O^{2-}_{(3\pm\gamma)}$$

Here, $x$ is the amount of $Mn^{4+}$ in the total Mn present in the sample.

Since, through titration experiment the value of $\gamma$ is found to be 0.042, therefore, according to charge neutrality condition

$$La^{3+}Ga^{3+}_{0.4}Mn^{3+}_{0.6(1-x)}Mn^{4+}_{0.6x}O^{2-}_{(3+\gamma)} = La^{3+}Ga^{3+}_{0.4}Mn^{3+}_{0.6(1-x)}Mn^{4+}_{0.6x}O^{2-}_{(3+0.042)} = 0$$

$(3 \times 1) + (3 \times 0.4) + [3 \times 0.6(1-x)] + (4 \times 0.6x) + (-2 \times 3.042) = 0$

$3 + 1.2 + 1.8 - 1.8x + 2.4x - 6.084 = 0$

$6 + 0.6x - 6.084 = 0$

$0.6x - 0.084 = 0$

$x = \dfrac{0.084}{0.6}$

$x = 0.14$

Thus $Mn^{4+}$ is 14% out of total Mn present in the sample whereas remaining 86% is $Mn^{3+}$. In this way, the charge state of Mn and percentage of $Mn^{3+}$ and $Mn^{4+}$ has been calculated. This estimated Mn charge state (3.14) (and hence the back calculated $\gamma$ and percentage of $Mn^{3+}$ and $Mn^{4+}$) was found closely consistent with the one determined through XANES measurements performed at Mn K-edge energy of 6539 eV.

**(2)  Room-temperature loss tangent (tanδ) as a function of frequency**

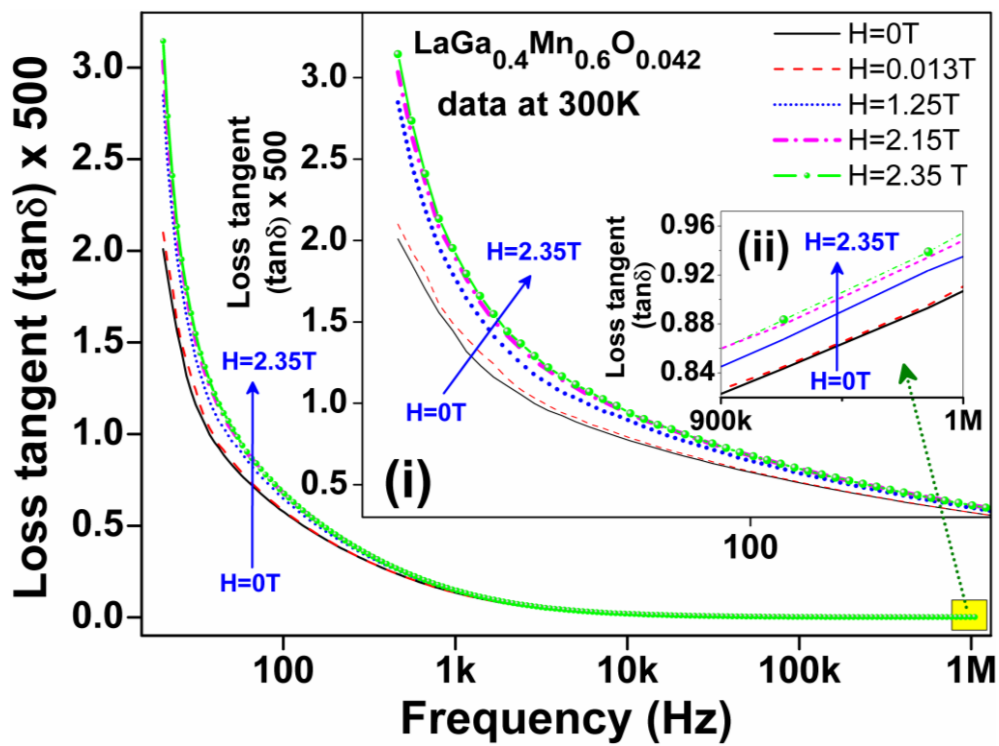

Figure-S1: Room-temperature loss tangent (tanδ) as a function of frequency (measured in the absence (H=0 T) and presence (upto H=2.35 T) of magnetic field) for LGM6O. Insets (i) and (ii) shows the magnified view of data across the corresponding range of probing frequencies.

*Note-* From application point of view, high loss tangent of $LaGa_{0.4}Mn_{0.6}O_{3.042}$ can be a factor of concern as it can make difficult for this material to be used in microelectronic devices because such devices require a material with very low *tanδ* (low leakage). In $LaGa_{0.4}Mn_{0.6}O_{3.042}$, as the leakage is associated with the coexistence of $Mn^{3+}$ and $Mn^{4+}$, the value of *tanδ* can be minimized by making the samples with better oxygen stoichiometry (i.e. with $Mn^{3+}$ alone) by means of annealing of the sample in a proper gas environment under precisely controlled and calibrated gas flow.